\documentclass[journal]{IEEEtran}
\usepackage[T1]{fontenc}
\usepackage{amsmath,amssymb,amsfonts}
\usepackage{graphicx}
\usepackage{booktabs}
\usepackage{enumitem}
\usepackage{float}
\usepackage{placeins}   
\usepackage{tikz}
\usetikzlibrary{shapes.geometric,arrows.meta,positioning,fit,backgrounds}
\usepackage{cite}
\usepackage{microtype}
\usepackage{tabularx}
\usepackage{multirow}
\usepackage{array}
\usepackage{mathtools}
\usepackage{balance}
\usepackage{hyperref}
\hypersetup{colorlinks=false}
\newcommand{\Heff}{H_\mathrm{eff}}
\newcommand{\HE}{H_\mathrm{energy}}
\newcommand{\HP}{H_\mathrm{power}}
\newcommand{\HC}{H_\mathrm{control}}
\newcommand{\Hcap}{H_\mathrm{cap}}
\newcommand{\Hset}{H_\mathrm{set}}

\newcommand{\Hphys}{H_\mathrm{phys}}
\newcommand{\Hvirt}{H_\mathrm{virt}}
\newcommand{\tact}{\tau_\mathrm{act}}
\newcommand{\tma}{t_\mathrm{ma}}          
\newcommand{\Srat}{S_\mathrm{rated}}
\newcommand{\lambdaMPP}{\lambda_\mathrm{MPP}}
\newcommand{\dotf}{|\dot{f}|_\mathrm{max}}
\title{On the Achievable Inertia Constant of\\
Inverter-Based Resources}
\author{Christos M. Nikolakakos,
        Hassan Haes~Alhelou,~\IEEEmembership{Senior Member,~IEEE,}
        and~Nikos~Hatziargyriou,~\IEEEmembership{Life~Fellow,~IEEE}%
\thanks{C. M. Nikolakakos and N. Hatziargyriou are with the School of Electrical and Computer Engineering, National Technical University of Athens, Athens, Greece (e-mail: cnikolakakos@mail.ntua.gr, nh@power.ece.ntua.gr).}%
\thanks{H.H. Alhelou is with the School of Engineering, Massachusetts Institute of Technology (MIT), Cambridge, MA 02139, USA (e-mail: alhelou@ieee.org).}%
}
\begin{document}
\maketitle
\begin{abstract}
As inverter-based resources (IBRs) displace synchronous generators,
the inertia they can actually contribute to the grid becomes a critical
planning parameter. Unlike synchronous machines, this contribution is
bounded simultaneously by the available energy reserve, the converter
power rating combined with voltage ride-through (VRT) obligations,
and the inertia-emulation control scheme with its activation delay.
This paper derives each bound in closed form and combines them into a
unified envelope $\Heff(t,\lambda,V_g)=\min(\HE,\HP,\HC)$, which uses
parameters accessible to the system operator to quantify how much
inertia a plant can provide at a given loading $\lambda$, grid voltage
$V_g$, and time $t$ after a disturbance.
The analysis shows that below a loading-dependent critical voltage,
VRT reactive-current priority does not leave active-current headroom
for inertial power injection; that the converter overload ratio
$\kappa$ matters mainly during voltage dips; and that de-loading
yields a linear, quantifiable inertia gain, while collocated
energy storage (ESS) removes the source-coupling constraint, making
grid-following (GFL) and grid-forming (GFM) plants with storage
equivalent post-activation, up to the GFL activation-delay discount
within short RoCoF measurement windows.
Simulations on the IEEE 9-bus system confirm
the framework and motivate a two-dimensional inertia capability curve
$H(\lambda,V_{g,\min})$ as a grid-code instrument analogous to the
P--Q diagram of synchronous generators.
\end{abstract}
\section{Introduction}
\IEEEPARstart{T}{he} displacement of synchronous generators (SGs)
by inverter-based resources (IBRs)---photovoltaic (PV) plants,
wind turbines, and battery energy storage systems (BESS)---is
fundamentally altering the inertial response of power systems.
Milano~\emph{et~al.}\ \cite{Milano2018} identify low inertia as a
foundational challenge for future grids, and Tielens and Van~Hertem
\cite{Tielens2016} quantify the link between inertia decline and
frequency security margins.
In the Continental European synchronous zone, system inertia has declined
by approximately 30\% over the past decade \cite{EntsoE2020}, while several
island systems already operate with near-zero synchronous inertia \cite{IrelandEirGrid2020}.
The resulting operational challenges are higher rates of change of
frequency (RoCoF), lower frequency nadirs, and reduced time for
operator intervention; the underlying frequency-response fundamentals
are covered in the classical references \cite{Kundur1994,Anderson1990}.
IBR control strategies can emulate inertial behavior: grid-forming
(GFM) virtual synchronous machines (VSM), pioneered by Beck and Hesse
\cite{Beck2007} and extended to synchronverter \cite{Zhong2011} and
distributed VSM \cite{DArco2015} architectures, and grid-following
(GFL) $\mathrm{d}f/\mathrm{d}t$ loops \cite{Dreidy2017}.
However, the inertia that such plants can actually
provide is constrained by three factors that are not captured by the
nameplate value programmed into the controller:
\begin{enumerate}[leftmargin=*, label=\arabic*)]
  \item the energy available on the DC side (capacitor, collocated storage, or de-loaded power headroom);
  \item the converter overcurrent rating and its interaction with VRT reactive-current obligations; and
  \item the inertia-emulation control scheme together with its activation delay.
\end{enumerate}
\subsection{Literature Gap}
Prior works address the three constraints in isolation.
Studies focused on the available energy reserve (e.g.\ DC-link
sizing, battery dispatch, and de-loading strategies) treat the
converter as an unconstrained current source.
Fang~\emph{et~al.}\ \cite{Fang2018} derive the DC-link inertia constant
from first principles without considering the current-rating constraint;
Zhu~\emph{et~al.}\ \cite{Zhu2013} apply a similar energy-based approach
to VSC-HVDC inertia emulation.
The review in \cite{FangReview2019} surveys virtual-inertia schemes at
fixed loading and nominal voltage, so the dependence on operating
condition is not captured.
Eriksson~\emph{et~al.}\ \cite{Eriksson2018} distinguish synthetic inertia
from fast frequency response without quantifying the operating-state
dependence of either service.
Control-oriented studies design or characterize GFM and GFL emulation
schemes assuming unlimited energy and power headroom:
D'Arco~\emph{et~al.}\ \cite{DArco2015} and Zhong and Weiss
\cite{Zhong2011} develop VSM control structures,
Bevrani~\emph{et~al.}\ \cite{Bevrani2014} survey virtual synchronous
generator concepts, and Li~\emph{et~al.}\ \cite{Li2022} establish a
duality between GFM and GFL paradigms---all at nominal operating conditions.
Poolla~\emph{et~al.}\ \cite{Poolla2019} derive optimal virtual-inertia
placement but model each IBR as a lossless injector with no current limit.
Studies of converter hardware ratings \cite{Rocabert2012}
treat overcurrent capability and fault ride-through separately from
inertia delivery.
Inertia estimation studies---including the PMU-based method of
Wall and Terzija \cite{Wall2014}, the multi-timescale approach of
Cai~\emph{et~al.}\ \cite{Cai2020}, and the data-driven IBR-specific
estimator of Tan and Zhao \cite{Tan2023}---develop online tools but
do not model the physical bounds that constrain the quantity they
estimate.
Because these constraints have mostly been studied separately, the
limiting factor for a given operating condition is not always clear.
Moreover, the limiting factor can change during a disturbance: a plant
that can provide inertial active power at nominal voltage may lose that
capability when VRT reactive current is prioritized. Existing
methods therefore do not provide a direct map from plant operating
condition to achievable IBR inertia, characterized by parameters
available to the system operator.
\subsection{Contributions}
The main contributions of this paper are:
\begin{enumerate}[leftmargin=*, label=\arabic*)]
  \item A unified inertia envelope
    $\Heff(t,\lambda,V_g)=\min(\HE,\HP,\HC)$ that maps achievable
    IBR inertia across the operating space using TSO-accessible
    parameters (Section~\ref{sec:framework}).
  \item Identification of a critical grid voltage
    $V_g^*(\lambda,\kappa,K_q)$ below which VRT reactive-current
    priority does not leave active-current headroom for inertial power
    injection (Section~\ref{sec:power}).
  \item Analytical proof that the overload ratio $\kappa$ does not
    affect achievable inertia at normal grid voltage in plants without
    collocated ESS, becoming relevant only during voltage dips
    through VRT current re-allocation (Section~\ref{sec:power}).
  \item A closed-form expression linking the GFL activation delay
    $\tact$ to the TSO's RoCoF measurement window $\tma$ and the
    resulting apparent inertia $H_\mathrm{app}$ (Section~\ref{sec:control}).
  \item An IBR inertia capability curve $H(\lambda,V_{g,\min})$,
    analogous to the conventional P--Q diagram, proposed as an
    instrument for grid connection agreements (Section~\ref{sec:operational}).
\end{enumerate}
The remainder of the paper is organized as follows.
Section~\ref{sec:defs} defines the inertia-related quantities
used throughout.
Section~\ref{sec:framework} introduces the core $\min(\cdot)$
formulation.
Sections~\ref{sec:energy}--\ref{sec:control} derive each bound in detail.
Section~\ref{sec:envelope} assembles the complete inertia envelope and
connects it to the TSO's RoCoF measurement window.
Section~\ref{sec:simulation} presents numerical results on the
IEEE 9-bus system.
Section~\ref{sec:operational} discusses operational applications.
Section~\ref{sec:conclusions} concludes.
\section{Inertia Definitions}
\label{sec:defs}
Four inertia-related quantities are used throughout the paper.
\textbf{Physical inertia.}
For an SG, $\Hphys = \frac{1}{2}J\omega_0^2/\Srat$ is a fixed machine
parameter. The rotor responds instantaneously to a torque imbalance,
without control action or activation delay.
\textbf{Virtual inertia.}
Emulated inertial response implemented via a virtual swing equation or
VSM control loop \cite{DArco2015,Bevrani2014}. A GFM inverter emulates
a rotating mass with constant $\Hvirt$, backed by stored energy, and
its response is activated at $\tact \approx 0$. $\Hvirt$ is subject to
energy, power, and control constraints; it is not a fixed physical
property.\\
\textbf{Achievable (capability) inertia.}
The inertia a plant can actually deliver, bounded by its energy,
power, and control constraints. Its time-dependent value is the
envelope $\Heff(t,\lambda,V_g)$ derived in
Section~\ref{sec:framework}; its post-activation value is denoted
$\Hcap(\lambda,V_g) = \Heff(t \geq \tact,\lambda,V_g)$.\\
\textbf{Apparent (measured) inertia.}
The inertia $H_\mathrm{app}$ perceived by the system operator based on
frequency measurements \cite{Tan2023,Dreidy2017}. These measurements are
estimated over a RoCoF measurement window of duration $\tma$
prescribed by the applicable grid operating code---for example, the
Frequency Operating Standard (FOS) of AEMO \cite{AEMC_FOS2023}---and
delivered through SCADA or PMU telemetry.
Because this window smooths the frequency trajectory over a time
interval that may include the GFL activation ramp, the resulting
estimate can differ substantially from the achievable inertia
$\Hcap$; in general $H_\mathrm{app} \neq \Hcap$, and the gap depends
systematically on $\tma$ and $\tact$ (Section~\ref{sec:control}).
These definitions are used below: for SGs, $\Hphys = \Hcap$ (instantaneous, no artifact);
for GFMs, $\Hcap \leq \Hvirt$ and $H_\mathrm{app} \approx \Hcap$;
for GFLs, $\Heff(t) \leq K_\mathrm{inertia} \cdot \min(1,\,t/\tact)$
and $H_\mathrm{app}$ depends strongly on $\tma/\tact$.
\begin{figure}[!t]
\centering
\resizebox{\columnwidth}{!}{%
\begin{tikzpicture}[
    node distance=0.6cm,
    box/.style={draw, rounded corners, text width=4.8cm, minimum height=1.0cm,
                align=center, font=\small},
    sbox/.style={draw, rounded corners, text width=3.0cm, minimum height=0.8cm,
                 align=center, font=\small\itshape},
    arr/.style={-{Stealth[length=2.2mm]}, thick},
]
\node[draw, rounded corners, fill=red!10, text width=8.5cm, align=center,
      font=\small] (meas)
     {$H_\mathrm{app} = g\!\bigl(\Heff,\;\tma,\;\text{RoCoF measurement method}\bigr)$};
\node[above=0.08cm of meas, font=\small\bfseries] {Measurement / Estimation Layer};
\node[box, fill=blue!15, below=1.0cm of meas] (core)
     {$\Heff = \min\bigl(\HE,\;\HP,\;\HC\bigr)$};
\draw[arr] (core) -- (meas);
\node[box, fill=yellow!20, right=2.0cm of core] (sg)
     {$H_\mathrm{SG} = \dfrac{\frac{1}{2}J\omega_0^2}{\Srat}$\\[3pt]
      Constant, physics-driven};
\node[above=0.06cm of sg, font=\small\bfseries] {SG Reference};
\draw[arr, dashed] (core) -- node[above,font=\footnotesize]{vs.} (sg);
\node[sbox, fill=green!15, below left=1.0cm and 1.8cm of core] (E)
     {\textbf{Energy Bound} $\HE$};
\node[sbox, fill=orange!15, below=1.0cm of core] (P)
     {\textbf{Power Bound} \\ $\HP$};
\node[sbox, fill=cyan!15, below right=1.0cm and 1.8cm of core] (C)
     {\textbf{Control Bound} $\HC$};
\draw[arr] (E) -- (core);
\draw[arr] (P) -- (core);
\draw[arr] (C) -- (core);
\node[font=\scriptsize, below=0.12cm of E, text width=3.2cm, align=center]
     {DC-link capacitor energy\\Collocated ESS energy\\De-loaded / available headroom};
\node[font=\scriptsize, below=0.12cm of P, text width=3.2cm, align=center]
     {Overload ratio $\kappa$\\Loading $\lambda$\\VRT: $I_p$--$I_q$ trade-off};
\node[font=\scriptsize, below=0.12cm of C, text width=3.2cm, align=center]
     {GFM: $\Hset$, $\tact \approx 0$\\GFL: ramp $\propto t/\tact$};
\end{tikzpicture}%
}
\caption{Unified framework for achievable IBR inertia: $\Heff$ as the
  minimum of energy, power, and control bounds.}
\label{fig:concept}
\end{figure}
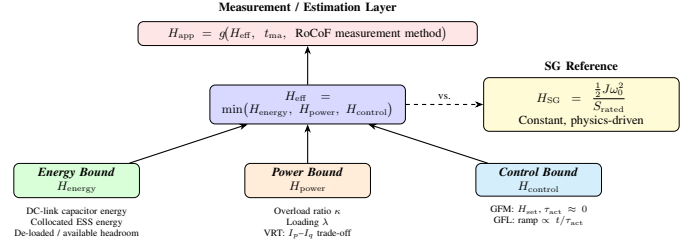
\section{Operating-State Dependence of IBR Inertia}
\label{sec:framework}
The effective inertia constant of an IBR is bounded by the minimum
of three coupled limits:
\begin{multline}
  \Heff(t, \lambda, V_g) = \\
  \min\!\bigl(\HE(t,\lambda),\;
  \HP(\lambda,\kappa,V_g),\;
  \HC(t)\bigr)
  \label{eq:core}
\end{multline}
where $t$ is the time elapsed since the onset of the disturbance,
$\lambda = P_\mathrm{op}/\Srat$ is the pre-disturbance loading, and
$V_g$ is the grid voltage at the point of common coupling after the
disturbance (in pu of $V_\mathrm{nom}$).
Fig.~\ref{fig:concept} illustrates the architecture;
the following sections derive each bound.
For an SG, all three bounds reduce to a single physical constant,
$H_\mathrm{SG} = \frac{1}{2}J\omega_0^2/\Srat$, independent of
$\lambda$, $V_g$, and $t$. This dependence on loading, voltage, and
time is the key distinction between SG inertia and achievable IBR
inertia.
\section{Energy Bound --- $\HE$}
\label{sec:energy}
\subsection{DC-Link Capacitor}
A plant of $N$ parallel modules (each rated $S_i$) has
aggregate DC-link energy given by the difference of the capacitor
stored energy $\tfrac{1}{2}CV^2$ between the upper and lower
admissible DC voltages:
\begin{equation}
  E_\mathrm{dc} = \tfrac{1}{2}C_\mathrm{dc,tot}
    \!\left(V_{\mathrm{dc,max}}^2 - V_{\mathrm{dc,min}}^2\right)
  \label{eq:Edc}
\end{equation}
with $C_\mathrm{dc,tot} = N C_{\mathrm{dc},i}$.
$V_\mathrm{dc,max}$ and $V_\mathrm{dc,min}$ are set by the capacitor
ripple/transient window specified in the converter datasheet and the
minimum DC voltage required to maintain AC modulation; the usable
fraction $(V_\mathrm{dc,max}^2 - V_\mathrm{dc,min}^2)$ is therefore
proportional to the energy the converter can release into the inertial
reference frame without loss of control \cite{Fang2018}.
Since $\Srat = N S_i$, the per-unit energy bound is:
\begin{equation}
  \HE^\mathrm{dc}
  = \frac{C_{\mathrm{dc},i}\!\left(V_{\mathrm{dc,max}}^2 - V_{\mathrm{dc,min}}^2\right)}{2\,S_i}
  \label{eq:Hdc}
\end{equation}
This is independent of $N$; only the per-module ratio
$C_{\mathrm{dc},i}/S_i$ determines $\HE^\mathrm{dc}$
\cite{Fang2018}.
For a typical utility-scale module ($C_{\mathrm{dc},i}\sim 1$\,mF,
$S_i\sim 0.5$\,MVA, $V_\mathrm{dc}\approx 1000$\,V,
$\Delta V_\mathrm{dc}\approx\pm10\%$), we obtain
$\HE^\mathrm{dc}\approx 0.4$\,ms---orders of magnitude lower than the
SG inertia.
\subsection{Collocated Battery Energy Storage System}
\begin{equation}
  \HE^\mathrm{ESS}
  = \frac{E_\mathrm{nom}\cdot\Delta\mathrm{SoC}\cdot\eta_\mathrm{dc-dc}}{\Srat}
  \label{eq:Hess}
\end{equation}
where $\Delta\mathrm{SoC} = \mathrm{SoC}_\mathrm{max} - \mathrm{SoC}_\mathrm{min}$
and $\eta_\mathrm{dc-dc}$ is converter efficiency.
For a plant with collocated BESS, $\HE^\mathrm{ESS}$ is typically large
(tens of seconds to minutes), making the energy bound non-binding over
the durations at which inertial support is actually used (of the order
of seconds) \cite{FangReview2019}.
$\HE^\mathrm{ESS}$ is SoC-dependent: for a 50\,MWh (180\,GJ) BESS on a
100\,MVA plant ($\eta_\mathrm{dc-dc} = 0.95$),
$\HE^\mathrm{ESS} = 1{,}710$\,s at full SoC, decreasing linearly to
$1{,}368$\,s at $\Delta\mathrm{SoC} = 0.8$ and to zero at the minimum
SoC threshold.
\begin{figure}[!htbp]
  \centering
  \includegraphics[width=\columnwidth]{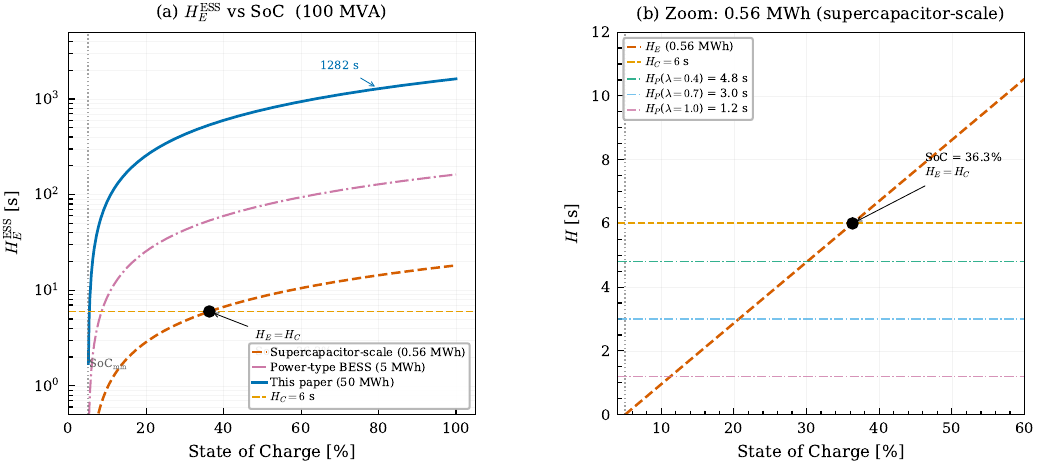}
  \caption{SoC-dependent energy bound ($\eta = 0.95$, 100\,MVA):
    (a)~$\HE^\mathrm{ESS}$ vs.\ SoC for three storage sizes;
    (b)~zoom on 0.56\,MWh showing crossover with $\HC$ and $\HP(\lambda)$.}
  \label{fig:soc_bound}
\end{figure}
Fig.~\ref{fig:soc_bound} illustrates this dependence for three
representative storage sizes on a 100\,MVA plant.
For the 50\,MWh BESS adopted in this paper, $\HE^\mathrm{ESS}$
ranges from 1{,}282\,s at $\mathrm{SoC} = 0.8$ to 1{,}624\,s at
full charge.
Even a 5\,MWh power-type battery yields $\HE = 128$\,s.
\subsection{De-Loaded / Curtailed Generation}
When a PV or wind plant operates below its maximum power point (MPP),
the primary source can provide additional active power
$\Delta P=(\lambdaMPP-\lambda)\Srat$ over the short time intervals
relevant to inertial support, provided the resource is available
\cite{FangReview2019}.
Therefore, for plants without collocated storage, the energy constraint
switches depending on the operating point:
\begin{equation}
  \HE(\lambda) =
  \begin{cases}
    \HE^\mathrm{dc} & \lambda = \lambdaMPP \\[3pt]
    +\infty\;\;(\text{non-binding})
      & \lambda < \lambdaMPP .
  \end{cases}
  \label{eq:He_deload}
\end{equation}
The ``$+\infty$'' entry does not imply that the plant contains
unbounded stored energy. It only means that, over the time scale on
which inertial support is used, the available power headroom rather
than the energy reserve becomes the relevant constraint. At MPP,
the plant has no primary-source headroom and the energy bound is set
by the DC-link capacitor. Below MPP, the binding constraint shifts to
the converter power limit $\HP$ and the control bound $\HC$.
Plants with collocated storage are treated separately in
Section~\ref{sec:gfl_ess}; in that case the source-coupling constraint
is removed and the energy bound is governed by the ESS energy reserve.
\section{Power / Rating Bound --- $\HP$}
\label{sec:power}
\subsection{Inverter Current Limit and Loading}
Define the inverter overload ratio $\kappa = S_\mathrm{max}/\Srat$,
i.e., the capability of the inverter to operate overloaded for a
short period of time.
Typical IBRs have $\kappa \approx 1.1$--$1.2$ \cite{Rocabert2012}.
For a plant emulating an inertia constant $H$, the stored kinetic
energy of the equivalent rotating mass is
$E = H\,\Srat\,(f/f_0)^2$; differentiating with respect to time and
evaluating at $f \approx f_0$ \cite{Kundur1994}, the inertial power
injection for a frequency event with RoCoF $\dot{f}$ is:
\begin{equation}
  P_\mathrm{inject}(t)
  = \frac{2H \cdot \Srat \cdot \dot{f}(t)}{f_0}
  \label{eq:Pinject}
\end{equation}
The constraint $P_\mathrm{inject}(t) \leq (\kappa - \lambda)\Srat$
limits the supportable inertia constant to:
\begin{equation}
  \HP = \frac{(\kappa - \lambda)\cdot f_0}{2\,\dotf}
  \label{eq:Hpower}
\end{equation}
At full load ($\lambda = 1.0$, $\kappa = 1.2$) under a
$|\dot{f}| = 1$\,Hz/s event at 50\,Hz, \eqref{eq:Hpower} gives
$\HP = 5$\,s.
Note that $\HP$ is defined relative to the design RoCoF $\dotf$ and
scales inversely with it, so the bound is specific to the assumed
design event.
This value is sufficient for the illustrative inertia setting used later
in this paper, but it can be reduced substantially by VRT obligations
and source--converter coupling, as shown below.
\subsection{Voltage Ride-Through Interaction}
During disturbances that produce a voltage depression, grid codes
prioritize reactive-current injection over active current
(e.g., the ENTSO-E RfG \cite{EntsoE2020}, NERC
fast-frequency-response guidelines \cite{NERC2020}, and IEEE~Std~2800
\cite{IEEE2800}).
The inverter current is constrained by the current circle:
\begin{equation}
  I_p^2 + I_q^2 \leq I_\mathrm{max}^2 = (\kappa I_\mathrm{nom})^2
  \label{eq:Icircle}
\end{equation}
Under a voltage dip below the VRT activation threshold $V_\mathrm{th}$
(the dead-band threshold per the applicable grid code
\cite{EentsoEgridcode2022,IEEE2800}, typically 0.9\,pu), the requested
VRT reactive current is:
\begin{equation}
  I_{q,\mathrm{VRT}} = \begin{cases}
    K_q \dfrac{V_\mathrm{th} - V_g}{V_\mathrm{nom}}\,I_\mathrm{nom}
      & V_g < V_\mathrm{th} \\[6pt]
    0 & V_g \geq V_\mathrm{th}
  \end{cases}
  \label{eq:IqVRT}
\end{equation}
where $K_q$ is the reactive-current droop and $V_g$ is the voltage at
the point of injection.
The residual active-current capacity is:
\begin{equation}
  I_{p,\mathrm{avail}}(V_g)
  = \sqrt{I_\mathrm{max}^2 - I_{q,\mathrm{VRT}}^2(V_g)}
  \label{eq:Ipavail}
\end{equation}
yielding a voltage-dependent effective overload ratio in power space:
\begin{equation}
  \kappa_\mathrm{eff}(V_g)
  = \frac{V_g \, I_{p,\mathrm{avail}}(V_g)}{V_\mathrm{nom}\, I_\mathrm{nom}}
  = \frac{V_g}{V_\mathrm{nom}}
    \sqrt{\kappa^2 - \!\left(\frac{K_q(V_\mathrm{th} - V_g)}{V_\mathrm{nom}}\right)^{\!2}}
  \label{eq:kappaeff}
\end{equation}
Substituting into \eqref{eq:Hpower}:
\begin{equation}
  \HP^\mathrm{VRT}(V_g, \lambda)
  = \frac{\bigl(\kappa_\mathrm{eff}(V_g) - \lambda\bigr)\cdot f_0}
         {2\,\dotf}
  \label{eq:HpowerVRT}
\end{equation}
It follows from \eqref{eq:HpowerVRT} that, under a disturbance with a
voltage depression $V_g \leq V_g^*$, where $V_g^*$ satisfies
$\kappa_\mathrm{eff}(V_g^*) = \lambda$, the power-limited inertia is
zero; i.e., the inverter has no active-current headroom for inertial
power injection while meeting its VRT reactive-current obligation.
For standard parameters $\kappa = 1.2$, $K_q = 2$, $V_\mathrm{th} = 0.9$\,pu:
at full load ($\lambda = 1.0$), $V_g^* = 0.84$\,pu;
at $\lambda = 0.85$, $V_g^* = 0.74$\,pu;
and at half load ($\lambda = 0.5$), $V_g^* = 0.53$\,pu.
$V_g^*$ is the grid voltage at which the VRT
reactive-current request exhausts the total active-current capacity
for a fixed loading
$\lambda$; it is \emph{not} the voltage the grid is predicted to reach
during the disturbance, which is determined by network topology and by
the inverter's own current injection.
As $\lambda$ increases, $V_g^*$ increases, meaning that a less severe
voltage dip is sufficient to eliminate the available active-current
headroom. Conversely, at lower loading, the inverter has more active
current margin, so a deeper voltage dip is required before inertial
active-power injection is blocked.
The practical implication is that a more heavily loaded inverter is
more vulnerable to losing inertial-response capability during voltage
depressions.
Table~\ref{tab:Hpower} illustrates $\HP^\mathrm{VRT}$ across scenarios
for $\kappa = 1.2$, $K_q = 2$, $\dotf = 1$\,Hz/s, $f_0 = 50$\,Hz.
\begin{table}[!htbp]
\renewcommand{\arraystretch}{1.15}
\caption{$\HP^\mathrm{VRT}$ across operating scenarios
         ($\kappa\!=\!1.2$, $K_q\!=\!2$, $\dotf\!=\!1$\,Hz/s)}
\label{tab:Hpower}
\centering
\footnotesize
\begin{tabular}{@{}lcccc@{}}
\toprule
Scenario & $\lambda$ & $V_g$ & $\kappa_\mathrm{eff}$ & $\HP$\,[s] \\
\midrule
no VRT    & 1.0 & 1.00 & 1.20 & 5.0 \\
mod.\ dip & 1.0 & 0.70 & 0.79 & $\approx$0 \\
sev.\ dip & 1.0 & 0.50 & 0.45 & $\approx$0 \\
no VRT    & 0.5 & 1.00 & 1.20 & 17.5 \\
mod.\ dip & 0.5 & 0.70 & 0.79 & 7.3 \\
\bottomrule
\end{tabular}
\end{table}
\subsection{Source--Converter Coupling (Plants without Collocated Storage)}
For plants without collocated ESS, the DC bus cannot buffer the
deficit between converter capability and primary-source output.
The sustained power headroom is:
\begin{equation}
  \frac{\Delta P_\mathrm{sus}}{\Srat}
  = \min\!\bigl(\kappa_\mathrm{eff}(V_g)\!-\!\lambda,\;\,
                \lambdaMPP\!-\!\lambda\bigr)
  \label{eq:Psus}
\end{equation}
giving a source-coupled power bound:
\begin{equation}
  \HP^\mathrm{no\,ESS}\!=\!
  \frac{f_0}{2\dotf}
  \min\!\bigl(\kappa_\mathrm{eff}(V_g)\!-\!\lambda,\;
              \lambdaMPP\!-\!\lambda\bigr)
  \label{eq:HpnoESS}
\end{equation}
For a plant without collocated ESS at $V_g \geq 0.9$\,pu,
\eqref{eq:kappaeff} gives $\kappa_\mathrm{eff} \approx \kappa > \lambdaMPP$,
so the minimum in \eqref{eq:HpnoESS} is determined by the source term
$(\lambdaMPP - \lambda)$ and is independent of $\kappa$.
Equivalently, plants with different overload ratios but the same
loading and no collocated storage deliver the same inertia at normal
voltage.
The overload ratio affects achievable inertia only during voltage
dips, through the VRT-induced reduction of $\kappa_\mathrm{eff}$ below
$\lambdaMPP$ via \eqref{eq:kappaeff}.
Consequently, procurement specifications that differentiate plants
only by $\kappa$ do not capture the inertia a plant can actually
provide at normal voltage.
\section{Control Bound --- $\HC$}
\label{sec:control}
\subsection{Grid-Forming Inverters (VSM / Droop / dVOC)}
A GFM inverter imposes its own voltage-frequency reference and
responds to torque imbalances equivalently---though not
identically---to an SG.
This behavior has been demonstrated across VSM \cite{DArco2015},
synchronverter \cite{Zhong2011}, and droop-based \cite{Denis2018}
architectures, and analyzed from duality \cite{Li2022} and placement
\cite{Poolla2019} perspectives.
The response is architecturally immediate:
\begin{equation}
  \HC^\mathrm{GFM}(t) = \Hset,
  \qquad \tact^\mathrm{GFM} \approx 0
  \label{eq:HcGFM}
\end{equation}
where $\Hset$ is the virtual inertia constant from the connection
agreement, verifiable through type-testing per
\cite{EentsoEgridcode2022,IEEE2800}.
Note that $\Hset$ is the \emph{commanded} value entering the control
bound, not the achievable inertia: the plant delivers
$\Heff \leq \Hset$, with equality only when the energy and power
bounds are non-binding.
Placement strategies for GFM virtual inertia to maximize system-level
benefit are examined in \cite{Poolla2019}.
\subsection{Grid-Following Inverters ($\mathrm{d}f/\mathrm{d}t$ Loop)}
A GFL inverter must first measure the grid frequency, compute
$\mathrm{d}f/\mathrm{d}t$, and only then command a power reference change.
The PLL-based frequency estimation introduces a finite activation
delay $\tact$ \cite{Dreidy2017,Fang2018}.
Modeling the build-up as a linear ramp---a first-order approximation
of the filter dynamics that uses TSO-accessible parameters:
\begin{equation}
  \HC^\mathrm{GFL}(t) = K_\mathrm{inertia}
  \cdot \min\!\!\left(1,\;\frac{t}{\tact^\mathrm{GFL}}\right)
  \label{eq:HcGFL}
\end{equation}
During the critical first 100--200\,ms after a disturbance---when the
RoCoF is typically steepest---\eqref{eq:HcGFL} gives
$\HC^\mathrm{GFL}(t) < K_\mathrm{inertia}$.
The effective inertia deficit during $0 \leq t < \tact$ is:
\begin{multline}
  \Delta H_\mathrm{deficit}(t) = K_\mathrm{inertia} - \HC^\mathrm{GFL}(t) \\
  = K_\mathrm{inertia}\!\left(1 - \frac{t}{\tact^\mathrm{GFL}}\right)
  \label{eq:deficit}
\end{multline}
This deficit is independent of loading and voltage; it is a
fundamental consequence of the GFL architecture.
A system operator relying on GFL synthetic inertia must discount
$K_\mathrm{inertia}$ when it is entered into the inertial-response
calculation.
\subsection{Link to the TSO's RoCoF Measurement Window}
\begin{figure}[!htbp]
  \centering
  \includegraphics[width=\columnwidth]{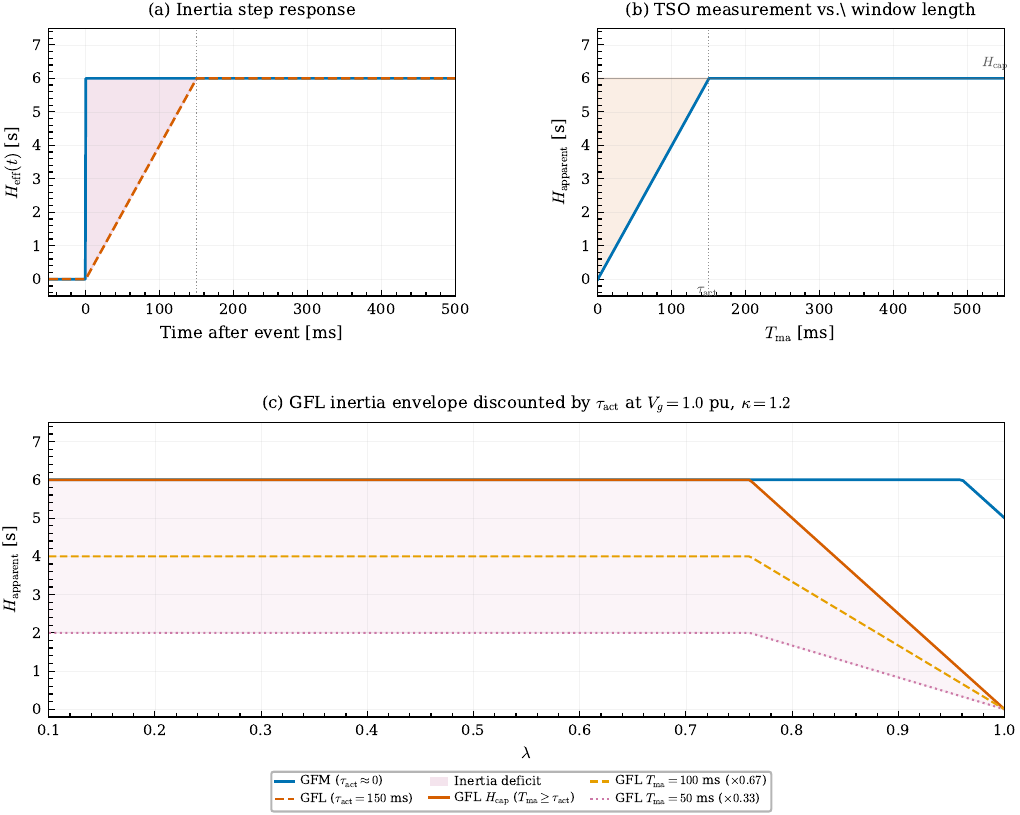}
  \caption{GFL activation delay impact ($\tact = 150$\,ms, $\kappa = 1.2$):
    (a)~$H_\mathrm{eff}(t)$ step response, GFM vs.\ GFL;
    (b)~apparent inertia $H_\mathrm{app}$ vs.\ TSO window $\tma$
    from \eqref{eq:Happarent};
    (c)~GFL envelope at $V_g = 1.0$\,pu discounted at $\tma
    = 50$ and $100$\,ms, compared to the GFM+ESS reference.}
  \label{fig:tact}
\end{figure}

Expression~\eqref{eq:HcGFL} provides a direct, quantitative link to
the window-length dependence of GFL inertia estimates.
A TSO estimating inertia from the instantaneous power at $t = \tma$
and the moving-average RoCoF over $[0,\,\tma]$ observes:
\begin{equation}
  H_\mathrm{app}^\mathrm{GFL}
  = \Hcap \cdot \min\!\!\left(\frac{\tma}{\tact},\; 1\right)
  \label{eq:Happarent}
\end{equation}
where $\Hcap = \min(\HE,\HP,\HC)$ is the achievable (capability)
inertia---the steady-state value once the converter is fully activated
(Section~\ref{sec:defs}).
\textit{Derivation.}
The GFL plant ramps linearly from zero to $P_\mathrm{del}$ over
$[0,\,\tact]$ and remains at $P_\mathrm{del}$ for $t > \tact$.
Assuming that the plant's injection does not appreciably alter the
system frequency trajectory, the system RoCoF remains approximately
constant at $\dot{\omega}_0$.
The moving-average RoCoF estimate
$\hat{\dot{\omega}} = [\omega(\tma) - \omega(0)] / \tma
\approx \dot{\omega}_0$.
The apparent inertia is then:
\begin{equation}
  H_\mathrm{app}
  = \frac{P(\tma)}{2\Srat}
    \cdot\frac{\omega_0}{\hat{\dot{\omega}}(\tma)}
  \approx \Hcap \cdot \frac{P(\tma)}{P_\mathrm{del}}
  \label{eq:Happ_def}
\end{equation}
where $\omega_0$ is the nominal angular frequency (not a derivative)
and $\hat{\dot{\omega}}$ is the moving-average RoCoF estimate above.
For $\tma \leq \tact$, the plant has not yet reached full output,
$P(\tma) = P_\mathrm{del} \cdot \tma/\tact$, giving
$H_\mathrm{app} = \Hcap \cdot \tma/\tact$.
For $\tma > \tact$, the plant is at full power,
$P(\tma) = P_\mathrm{del}$, giving $H_\mathrm{app} = \Hcap$.
The result is a sharp threshold: \emph{if the TSO window exceeds
$\tact$, the full $\Hcap$ is observed; otherwise the apparent inertia
scales linearly with $\tma/\tact$}.
For a GFL plant with $\Hcap = 6$\,s and $\tact = 150$\,ms,
\eqref{eq:Happarent} yields $H_\mathrm{app} = 2.0$\,s at $\tma = 50$\,ms
($0.33\times\Hcap$); $H_\mathrm{app} = 4.0$\,s at $\tma = 100$\,ms
($0.67\times\Hcap$); and $H_\mathrm{app} = \Hcap = 6.0$\,s for any
$\tma \geq 150$\,ms.
For GFM, $\tact \approx 0$ and $H_\mathrm{app} = \Hcap$ regardless of
$\tma$.
Fig.~\ref{fig:tact} visualizes this interaction.
Figure~\ref{fig:tact}(a) shows the $H_\mathrm{eff}(t)$ step response:
the GFM delivers $\Hcap$ immediately, while the GFL ramps linearly
from zero over $[0,\,\tact]$.
The shaded area is the cumulative inertia deficit---energy that the
system does not receive during the critical first cycles.
Figure~\ref{fig:tact}(b) plots $H_\mathrm{app}$ from
\eqref{eq:Happarent} as a function of $\tma$: the curve rises linearly
from zero, reaching $\Hcap$ at $\tma = \tact$ and remaining flat
thereafter.
Figure~\ref{fig:tact}(c) connects the $\tact$ penalty back to the
$(\lambda, V_g)$ inertia envelope by plotting the GFL cross-section at
$V_g = 1.0$\,pu for $\tma < \tact$.
At $\tma = 50$\,ms, the TSO-observed inertia is $0.33\times\Hcap$---the
GFL with $\Hcap = 6.0$\,s at $\lambda = 0.5$ appears as $2.0$\,s, less
than a typical SG.
At $\tma = 100$\,ms, it recovers to $0.67\times\Hcap$.
For any $\tma \geq \tact = 150$\,ms, the full post-activation envelope
is observed.
The GFM+ESS reference is unaffected by $\tact$ and represents the
true system contribution at all time scales.
The inertia envelopes presented in the remainder of this paper
correspond to the post-activation bound
$\Hcap(\lambda, V_g) = \min(\HE, \HP, \HC)$, i.e.\ assuming
$\tma \geq \tact$ for GFL plants.
\subsection{Measurement Window Impact: GFM+ESS vs.\ GFL+ESS}
\label{sec:meas_window}
Adding collocated ESS removes source coupling entirely: both GFL+ESS
and GFM+ESS deliver the same post-activation inertia
$\Hcap(\lambda, V_g) = \min(\HE, \HP, \HC)$.
However, the GFL activation delay $\tact$ introduces a measurement
asymmetry.
A TSO using a moving-average window $\tma < \tact$ observes only the
discounted value from \eqref{eq:Happarent}:
\begin{equation}
  H_\mathrm{app}^{\mathrm{GFL}} = \Hcap \times
    \min\!\left(\frac{\tma}{\tact},\;1\right),
  \label{eq:Happ_gfl}
\end{equation}
while the GFM response is instantaneous ($\tact \approx 0$), so
$H_\mathrm{app}^{\mathrm{GFM}} = \Hcap$ for any $\tma > 0$.
For the parameters used in this paper ($\tact = 150$\,ms,
$\tma = 100$\,ms), the discount factor is
$\tma/\tact = 0.667$: a GFL+ESS plant with $\Hcap = 6.0$\,s appears to
the TSO as $4.0$\,s, while an identical GFM+ESS plant receives full
credit.
\section{The Inertia Envelope}
\label{sec:envelope}
\subsection{Complete Formulation}
Combining Sections~\ref{sec:energy}--\ref{sec:control}:
\begin{multline}
  \Heff(t,\lambda,V_g) = \\
  \min\!\Bigl(\HE(t,\lambda),\;
    \HP\!\bigl(\lambda,\kappa,V_g\bigr),\;
    \HC(t)\Bigr)
  \label{eq:envelope}
\end{multline}
All inputs---$\lambda$, $V_g$, $\kappa$, $\lambdaMPP$, $E_\mathrm{ESS}$,
$\Hset$ or $K_\mathrm{inertia}$, and $\tact$---are available to the TSO
from SCADA, connection agreements, and type-test certificates.
No proprietary internal firmware parameters are required.
\subsection{GFM with Collocated ESS Example}
\begin{figure}[!htbp]
  \centering
  \includegraphics[width=\columnwidth]{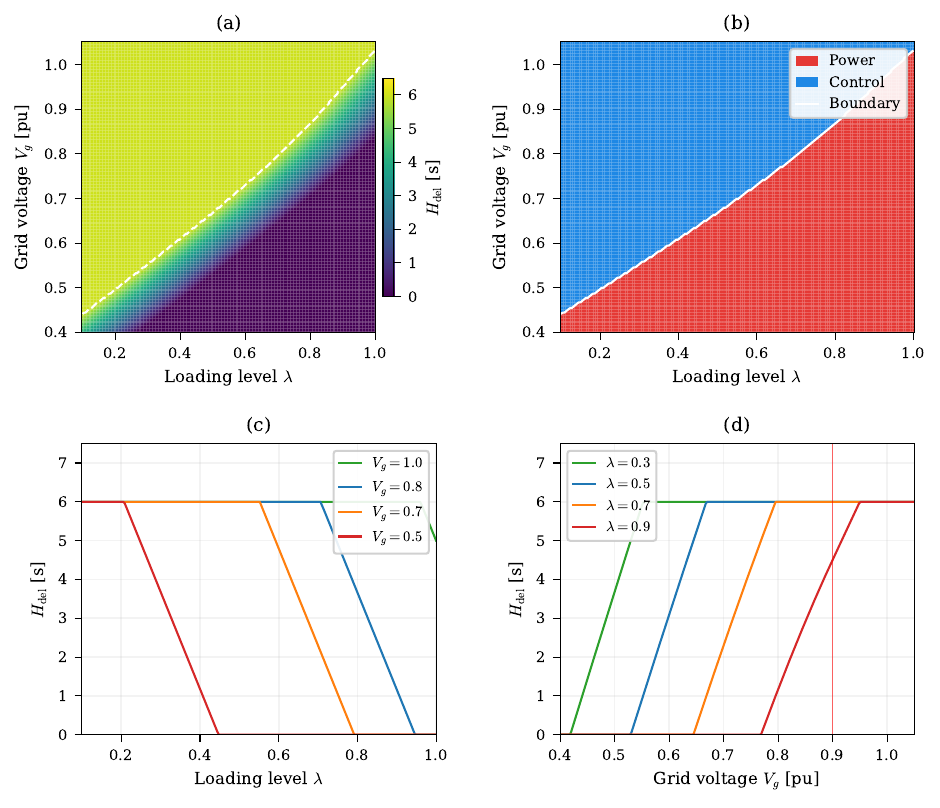}
  \caption{GFM + ESS inertia envelope (Table~\ref{tab:gfm_params}):
    (a)~$\Hcap(\lambda, V_g)$ magnitude,
    (b)~binding constraint map,
    (c)~loading slices,
    (d)~voltage slices.}
  \label{fig:gfm_envelope}
\end{figure}

\begin{figure}[!htbp]
  \centering
  \includegraphics[width=\columnwidth]{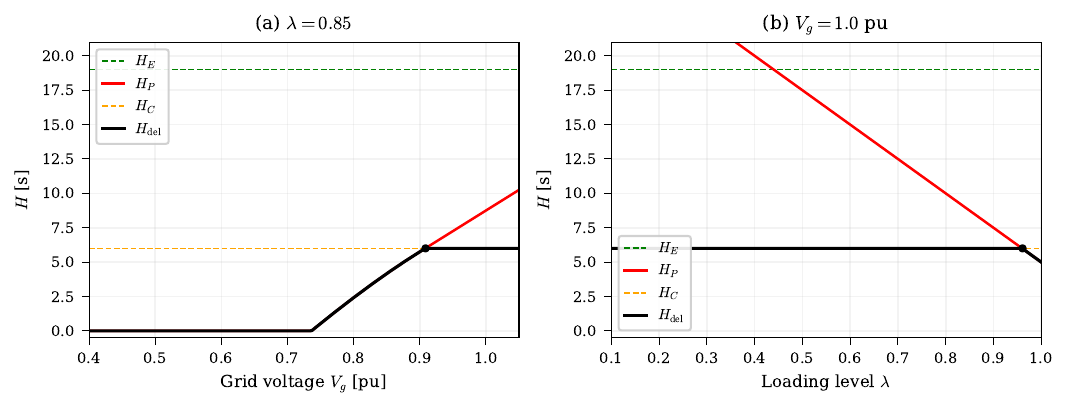}
  \caption{GFM bound decomposition:
    (a)~vs.\ grid voltage at $\lambda = 0.85$;
    (b)~vs.\ loading at $V_g = 1.0$\,pu.}
  \label{fig:gfm_decomp}
\end{figure}

For a 100\,MVA GFM inverter with BESS (parameters:
Table~\ref{tab:gfm_params}), the three bounds evaluate to:
$\HE \approx 1{,}710$\,s (non-binding),
$\HC = 6$\,s (constant, $\tact \approx 0$),
$\HP(\lambda,V_g)$ ranges from 17.5\,s ($\lambda = 0.5$, $V_g = 1.0$\,pu)
to 0 when $V_g$ falls below $V_g^*$
(e.g., $\HP = 0$ at $\lambda = 0.9$, $V_g \leq 0.78$\,pu).
The binding constraint is:
\begin{itemize}[leftmargin=*, noitemsep, topsep=2pt]
  \item $\HC = 6$\,s (control-limited) when $\HP > \HC$, i.e.\
    $(\kappa_\mathrm{eff}(V_g) - \lambda)\,f_0/2\dotf > 6$;
  \item $\HP$ (power-limited) otherwise, reducing to zero as VRT
    consumes all available current.
\end{itemize}
\begin{table}[!htbp]
\renewcommand{\arraystretch}{1.15}
\caption{Parameters: GFM Inertia Envelope (Section~\ref{sec:envelope})}
\label{tab:gfm_params}
\centering
\begin{tabular}{lll}
\toprule
Parameter & Symbol & Value \\
\midrule
Rated power       & $\Srat$        & 100\,MVA \\
Nom.\ frequency   & $f_0$          & 50\,Hz   \\
Overload ratio    & $\kappa$       & 1.2      \\
Virtual inertia   & $\Hset$        & 6\,s     \\
VRT gain          & $K_q$          & 2        \\
ESS energy        & $E_\mathrm{ESS}$ & 50\,MWh (180\,GJ) \\
Design RoCoF      & $\dotf$        & 1\,Hz/s  \\
\bottomrule
\end{tabular}
\end{table}
The resulting inertia envelope is shown in Fig.~\ref{fig:gfm_envelope}.
The bound-switching boundary between control-limited and power-limited
regimes is clearly visible: at $V_g = 1.0$\,pu the transition occurs at
$\lambda^* = \kappa_\mathrm{eff}(V_g) - 2\HC\dotf/f_0 = 0.96$; at
$V_g = 0.7$\,pu (moderate dip) it shifts to $\lambda^* = 0.55$; and at
$V_g = 0.5$\,pu (severe dip) to $\lambda^* = 0.21$.
Fig.~\ref{fig:gfm_decomp} decomposes the three bounds along
representative loading and voltage slices, confirming that
$\HC = 6$\,s is the binding constraint across most of the normal
operating region.
\subsection{GFL without Collocated ESS --- DC-Link Only}
\begin{figure}[!htbp]
  \centering
  \includegraphics[width=\columnwidth]{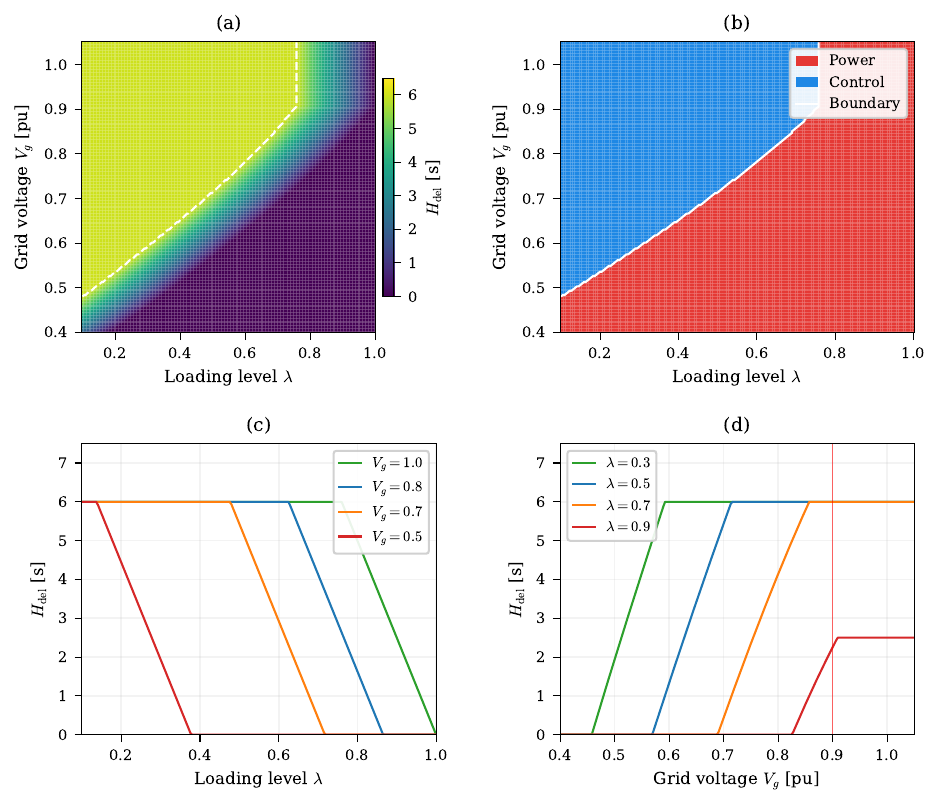}
  \caption{GFL de-loaded inertia envelope (Table~\ref{tab:gfl_params},
    $\lambda < \lambdaMPP$):
    (a)~$\Hcap(\lambda, V_g)$ magnitude,
    (b)~binding constraint map,
    (c)~loading slices,
    (d)~voltage slices.}
  \label{fig:gfl_envelope}
\end{figure}

\begin{figure}[!htbp]
  \centering
  \includegraphics[width=\columnwidth]{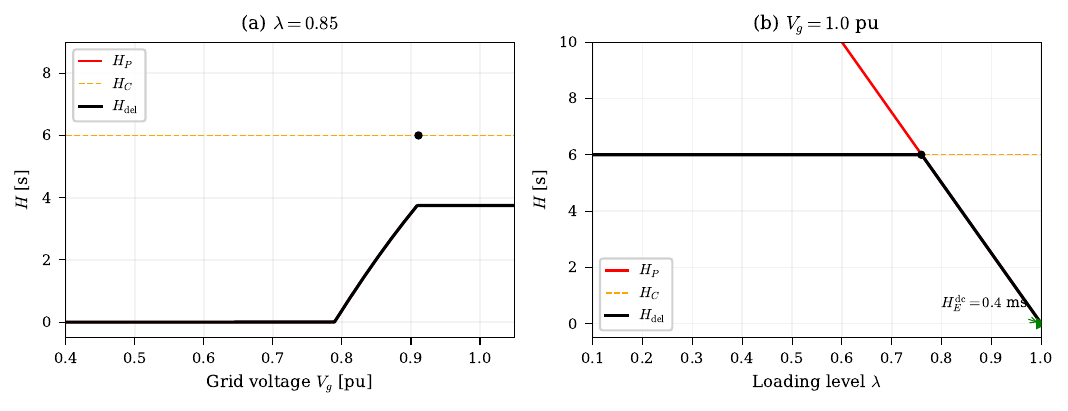}
  \caption{GFL bound decomposition:
    (a)~vs.\ grid voltage at $\lambda = 0.85$;
    (b)~vs.\ loading at $V_g = 1.0$\,pu.}
  \label{fig:gfl_decomp}
\end{figure}

For the GFL case (Table~\ref{tab:gfl_params}):
$\HE^\mathrm{dc} \approx 0.4$\,ms (energy-binding at MPP);
$\HC^\mathrm{GFL}(t) = 6\,$s at $t \geq \tact = 150$\,ms;
$\HP^\mathrm{no\,ESS}$ follows \eqref{eq:HpnoESS} with the source term
$(\lambdaMPP - \lambda)$ always binding at normal voltage.
The marginal inertia gain from de-loading is constant and exactly
quantifiable: each 1\% of available power headroom below MPP adds
\begin{equation}
  \Delta H = \frac{0.01\,f_0}{2\,\dotf} = 0.25\;\mathrm{s}
\end{equation}
evaluated at $f_0 = 50$\,Hz and $\dotf = 1$\,Hz/s.
At 24\% de-loading ($\lambda = 0.76$), the source-coupled $\HP$
equals the $K_\mathrm{inertia} = 6$\,s from the connection
agreement, at which point the control bound becomes binding and
further de-loading yields no additional inertia.
\begin{table}[!htbp]
\renewcommand{\arraystretch}{1.15}
\caption{Parameters: GFL without Collocated ESS (Section~\ref{sec:envelope})}
\label{tab:gfl_params}
\centering
\footnotesize
\begin{tabular}{@{}lll@{}}
\toprule
Parameter & Symbol & Value \\
\midrule
Module power     & $S_i$                   & 0.5\,MVA  \\
Parallel modules & $N$                     & 200       \\
Plant power      & $\Srat$                 & 100\,MVA  \\
DC-link cap.     & $C_{\mathrm{dc},i}$     & 1\,mF     \\
DC voltage       & $V_\mathrm{dc}$         & 1000\,V   \\
DC volt.\ tol.   & $\Delta V_\mathrm{dc}$  & $\pm10\%$ \\
Overload ratio   & $\kappa$                & 1.1       \\
Inertia gain     & $K_\mathrm{inertia}$    & 6\,s      \\
Activation time  & $\tact$                 & 150\,ms   \\
MPP loading      & $\lambdaMPP$            & 1.0       \\
\bottomrule
\end{tabular}
\end{table}
Fig.~\ref{fig:gfl_envelope} shows the resulting GFL inertia envelope
for de-loaded operation ($\lambda < \lambdaMPP$), where the source
sustains the headroom over the relevant intervals ($\HE = +\infty$).
The envelope exhibits a linear ramp in $\lambda$, saturating at
$\HC = 6$\,s for $\lambda \leq 0.76$ (24\% de-loading).
At MPP ($\lambda = 1.0$), the energy bound reduces to
$\HE^\mathrm{dc} = 0.4$\,ms and becomes binding everywhere,
regardless of voltage or control setting---a gap of approximately
$15{,}000\times$ relative to GFM+ESS.
Fig.~\ref{fig:gfl_decomp} decomposes the three bounds, highlighting
the source--converter coupling that limits headroom to
$(\lambdaMPP - \lambda)$ at normal voltage.
\subsection{GFL with Collocated ESS --- Decoupling the Source Constraint}
\label{sec:gfl_ess}
\begin{figure}[!htbp]
  \centering
  \includegraphics[width=\columnwidth]{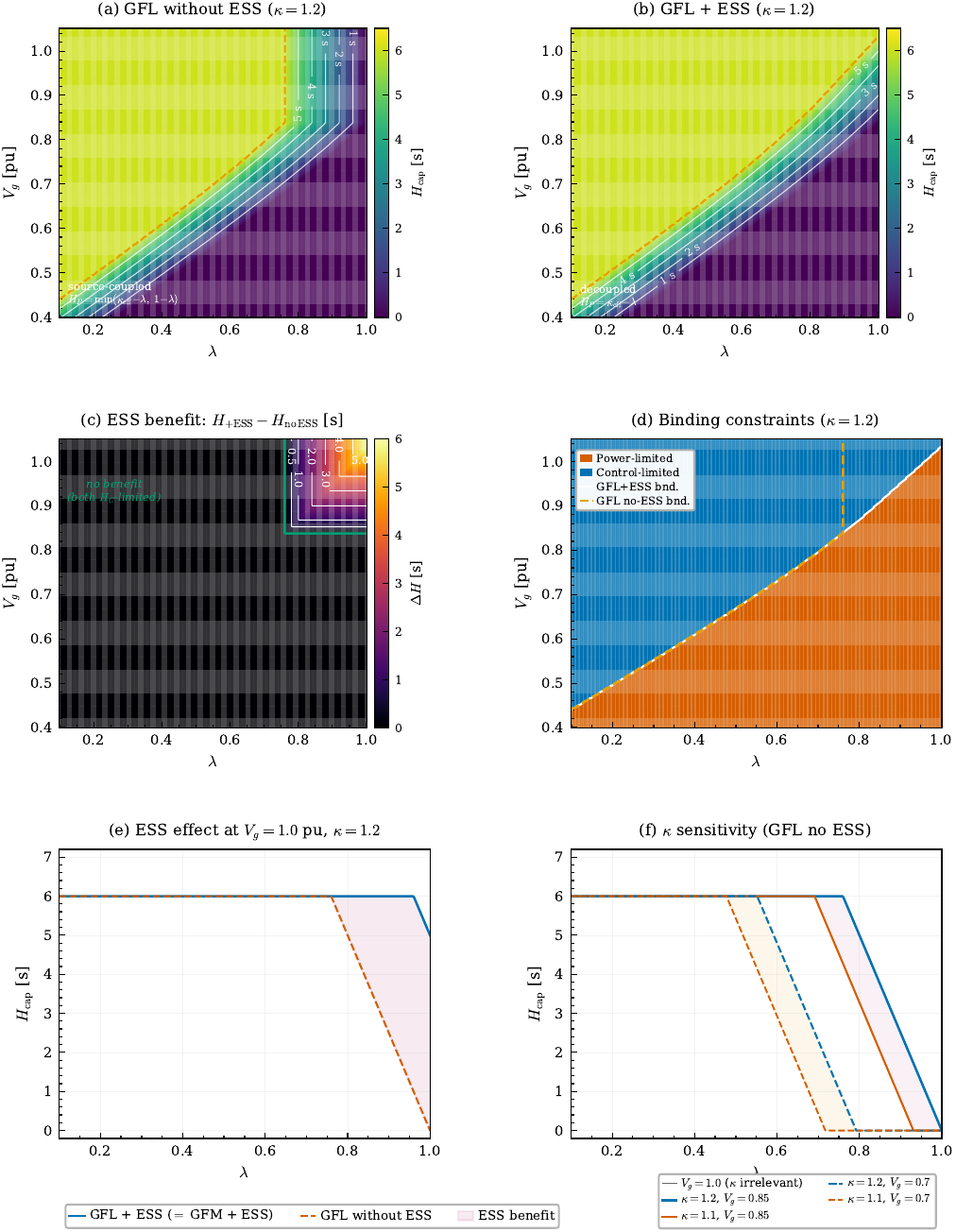}
  \caption{Effect of collocated ESS on the GFL inertia envelope
    ($\kappa = 1.2$):
    (a)~GFL without ESS; (b)~GFL with ESS ($\equiv$ GFM+ESS);
    (c)~ESS benefit $\Delta H$; (d)~binding constraint map;
    (e)~$\lambda$-sweep at $V_g = 1.0$\,pu;
    (f)~$\kappa$ sensitivity at $V_g = 0.85$ and $0.7$\,pu.}
  \label{fig:gfl_ess}
\end{figure}

The preceding analysis reveals that the GFL--GFM gap at normal
voltage originates primarily from source--converter coupling, which
caps the power headroom at $\min(\kappa_\mathrm{eff} - \lambda,\; 1 - \lambda)$
instead of $(\kappa_\mathrm{eff} - \lambda)$.
Adding a collocated BESS to the GFL plant removes this coupling---the
ESS delivers the incremental power independently of the primary-source
operating point.
The GFL+ESS envelope is therefore:
\begin{equation}
  \Hcap^\mathrm{GFL{+}ESS}(\lambda, V_g)
  = \min\!\bigl(\HE^\mathrm{ESS},\;
    \HP(\kappa, \lambda, V_g),\;
    \HC\bigr)
  \label{eq:gfl_ess}
\end{equation}
where $\HE^\mathrm{ESS} = 1{,}710$\,s (non-binding, identical to the
GFM case) and $\HP$ no longer contains the source term $(1 - \lambda)$.
At the same overload ratio $\kappa = 1.2$, the GFL+ESS and
GFM+ESS envelopes are \emph{identical}---the only structural difference
between GFL and GFM is the source coupling, which the collocated ESS
eliminates.
At $V_g = 1.0$\,pu the ESS benefit is quantified in
Table~\ref{tab:comparison}.
For $\lambda \leq 0.76$, both the GFL+ESS and the GFL without ESS
deliver $\Hcap = \HC = 6$\,s (control-limited); the ESS provides no
additional inertia in this regime.
Above $\lambda = 0.76$, the source-coupling constraint becomes binding
for the GFL without ESS while the GFL+ESS (= GFM+ESS) remains
control-limited up to $\lambda \approx 0.96$.
At $\lambda = 0.85$, the ESS adds $2.25$\,s (+60\%).
At $\lambda = 0.90$, the gain grows to $3.50$\,s, and at $\lambda = 0.95$
it reaches $4.75$\,s---recovering all of the GFM--GFL gap.
Although $\kappa$ is irrelevant at normal voltage for source-coupled
GFL (because $(1 - \lambda)$ always binds before $(\kappa - \lambda)$
when $\kappa > 1$), it becomes significant under VRT conditions.
Fig.~\ref{fig:gfl_ess}(f) shows that at $V_g = 0.85$\,pu the increase
from $\kappa = 1.1$ to $\kappa = 1.2$ improves the achievable inertia
appreciably, and the effect is amplified at $V_g = 0.7$\,pu where the
VRT reactive-current priority further reduces $\kappa_\mathrm{eff}$.
Fig.~\ref{fig:gfl_ess} presents the full $(\lambda, V_g)$ comparison,
including the shift of the power/control binding boundary when
collocated ESS is added.
At nominal or near-nominal voltage ($V_g \approx 1.0$\,pu), the
binding power headroom for a source-coupled GFL is $(1 - \lambda)$,
which vanishes as the plant approaches MPPT ($\lambda \to 1$).
With ESS the binding headroom becomes $(\kappa - \lambda)$, set by
the inverter overload capability rather than by the primary-source
operating point.
For $\kappa = 1.2$, the power--control boundary therefore shifts from
$\lambda \approx 0.76$ to $\lambda \approx 0.96$---an additional
$20$\,percentage points of the loading range over which the plant
delivers its full control-limited inertia $\Hcap = \HC$.
\subsection{Contrast with Synchronous Generators}
\begin{figure}[!htbp]
  \centering
  \includegraphics[width=\columnwidth]{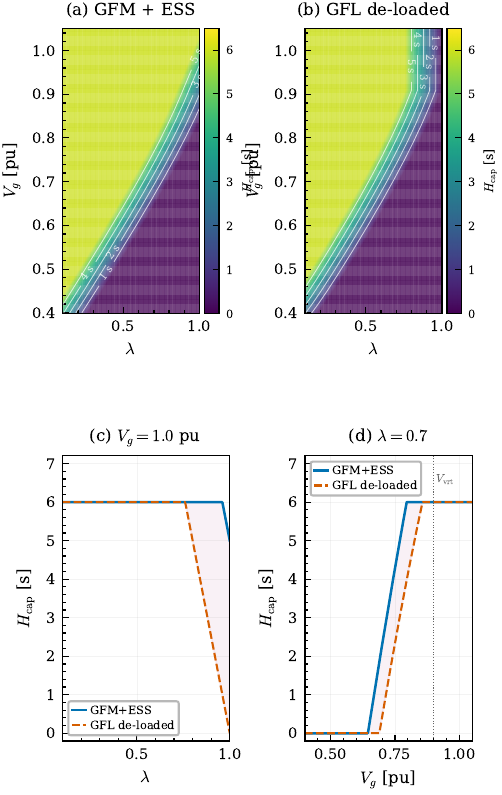}
  \caption{GFM+ESS ($\kappa = 1.2$) vs.\ GFL de-loaded ($\kappa = 1.1$):
    (a,\,b)~inertia envelopes;
    (c)~$\lambda$-sweep at $V_g = 1.0$\,pu;
    (d)~$V_g$-sweep at $\lambda = 0.7$.}
  \label{fig:comparison}
\end{figure}

For an SG with inertia constant $H_\mathrm{SG}$,
$\Heff^\mathrm{SG}(t,\lambda,V_g) = H_\mathrm{SG}\;\forall\;t,\lambda,V_g$.
For an IBR, $\Heff(t,\lambda,V_g)$ is a non-constant surface
in $(t,\lambda,V_g)$ space, equal to $H_\mathrm{SG}$ only in a
subset of the operating space and reducing to zero in a non-negligible
region.
The measure of the ``zero-inertia region'' (where $\Heff < \epsilon H_\mathrm{SG}$)
depends on the plant type.
For GFM+ESS, it is bounded by the VRT activation threshold.
For GFL without collocated ESS, it includes the entire MPP operating
line ($\lambda = \lambdaMPP$) and the initial part of the activation
ramp after any event.
For GFL+ESS at the same $\kappa$, the zero-inertia region is
identical to that of GFM+ESS---the collocated ESS entirely eliminates
the source-coupling penalty.
Table~\ref{tab:comparison} quantifies the comparison at normal
voltage ($V_g = 1.0$\,pu) across loading levels.
For $\lambda \leq 0.76$, both configurations deliver $\Hcap = 6$\,s
(control-limited).
Above $\lambda = 0.76$, the source-coupling constraint becomes
binding for the GFL without ESS while the GFL+ESS (= GFM+ESS)
remains at $6.0$\,s: at $\lambda = 0.90$, the ESS recovers $3.50$\,s,
and at $\lambda = 0.95$ it recovers $4.75$\,s.
\begin{table}[!htbp]
\renewcommand{\arraystretch}{1.15}
\caption{Delivered Inertia Comparison at $V_g = 1.0$\,pu}
\label{tab:comparison}
\centering
\begin{tabular}{cccc}
\toprule
$\lambda$ & GFL no ESS & GFL+ESS (= GFM+ESS) & ESS gain \\
          & $\Hcap$ [s] & $\Hcap$ [s] & [s] \\
\midrule
0.30 & 6.00 & 6.00 & 0.00 \\
0.50 & 6.00 & 6.00 & 0.00 \\
0.70 & 6.00 & 6.00 & 0.00 \\
0.80 & 5.00 & 6.00 & 1.00 \\
0.85 & 3.75 & 6.00 & 2.25 \\
0.90 & 2.50 & 6.00 & 3.50 \\
0.95 & 1.25 & 6.00 & 4.75 \\
1.00 & 0.00 & 5.00 & 5.00 \\
\bottomrule
\end{tabular}
\end{table}
Fig.~\ref{fig:comparison} visualizes the GFM--GFL contrast across the
full $(\lambda, V_g)$ operating space.
\subsection{Measurement Window Results}
\begin{figure}[!htbp]
  \centering
  \includegraphics[width=\columnwidth]{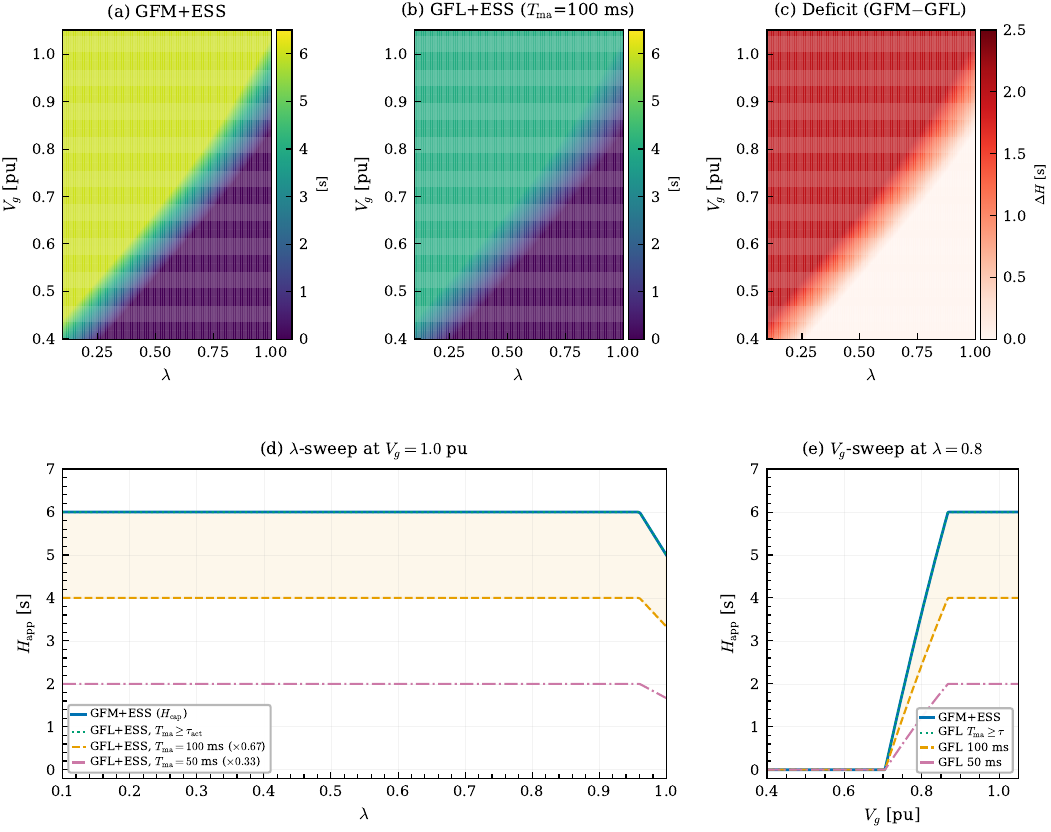}
  \caption{Measurement-window impact on apparent inertia
    ($\tact = 150$\,ms):
    (a)~GFM+ESS $H_\mathrm{app} = \Hcap$;
    (b)~GFL+ESS at $\tma = 100$\,ms;
    (c)~deficit $\Delta H$ (GFM$-$GFL);
    (d)~$\lambda$-sweep at $V_g = 1.0$\,pu;
    (e)~$V_g$-sweep at $\lambda = 0.8$.}
  \label{fig:meas_window}
\end{figure}

Fig.~\ref{fig:meas_window} quantifies the measurement-window asymmetry
across the full $(\lambda, V_g)$ operating plane.
Figure~\ref{fig:meas_window}(a) and (b) show the apparent-inertia
heatmaps for GFM+ESS (full $\Hcap$) and GFL+ESS ($0.667\times\Hcap$ at
$\tma = 100$\,ms), respectively.
Figure~\ref{fig:meas_window}(c) maps the inertia deficit
$\Delta H = H_\mathrm{app}^{\mathrm{GFM}} - H_\mathrm{app}^{\mathrm{GFL}}$,
peaking at ${\sim}2$\,s where $\Hcap = \HC$.
The $\lambda$-sweep at $V_g = 1.0$\,pu [Figure~\ref{fig:meas_window}(d)]
and $V_g$-sweep at $\lambda = 0.8$ [Figure~\ref{fig:meas_window}(e)]
compare GFM+ESS, GFL+ESS post-activation, and GFL+ESS at $\tma = 100$\,ms
and $50$\,ms.
The shaded region between the GFM and discounted GFL curves represents
the inertia that is physically achievable but invisible to the TSO.
At $\lambda = 0.8$ and $V_g = 1.0$\,pu, $\Hcap = 6.0$\,s for both
topologies, but the TSO observes only $4.0$\,s for GFL+ESS at
$\tma = 100$\,ms---a $2.0$\,s deficit.
At $\tma = 50$\,ms, this worsens to $4.0$\,s deficit
($H_\mathrm{app} = 2.0$\,s).
For GFL+ESS plants to receive equivalent TSO credit, either the
measurement window must satisfy $\tma \geq \tact$, or the plant must
oversize by a factor $\tact / \tma$ to compensate for the discount.
\section{Validation}
\label{sec:simulation}
Time-domain simulations are conducted in DIgSILENT PowerFactory on the IEEE 9-bus system comprising two
synchronous generators
(G1: $S_1 = 512$\,MVA, $H_1 = 2.63$\,s;
 G2: $S_2 = 270$\,MVA, $H_2 = 4.13$\,s)
and an IBR plant ($S_\mathrm{inv} = 125$\,MVA) at Bus~10.
Three loads are connected: Load~A (125\,MW), Load~B (90\,MW),
and Load~C (100\,MW).
In all cases, the triggering contingency is a 90\% step increase
in Load~A ($\Delta P = 112.5$\,MW); only the IBR configuration
(GFL or GFM) and its control settings are varied between scenarios.
\subsection{Impact of Collocated Energy Storage on GFL Inertia Contribution}
\label{sec:sim_ess}
\begin{figure}[!htbp]
  \centering
  \includegraphics[width=\columnwidth]{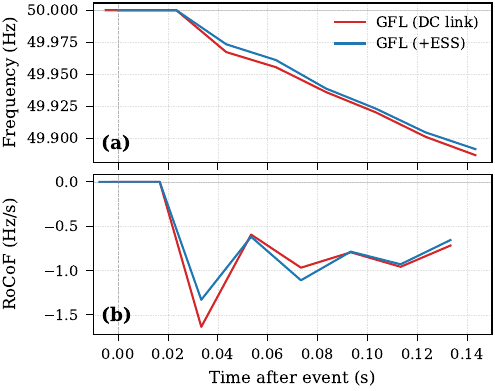}\\[4pt]
  \includegraphics[width=\columnwidth]{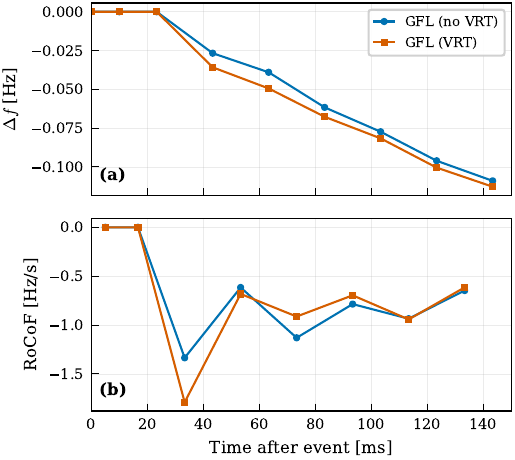}
  \caption{Time-domain validation on the IEEE 9-bus system
    (112.5\,MW load step).
    Top: (a)~frequency and (b)~unfiltered RoCoF for GFL with DC link
    only vs.\ GFL+ESS.
    Bottom: (a)~frequency deviation and (b)~unfiltered RoCoF with and
    without VRT activation ($V_g = 0.62$\,pu, $K_q = 5$).}
  \label{fig:sim_results}
\end{figure}

This test isolates the role of the available energy buffer by
comparing two otherwise identical GFL configurations:
(i)~GFL~(DC~link), where the inverter relies solely on its finite
DC-link capacitance ($\HE \approx 0.5$\,ms); and
(ii)~GFL~(+ESS), where a collocated battery energy storage system
makes $\HE$ non-binding.
Both cases share $\kappa = 1.2$ and $K_\mathrm{inertia} = 4.7$\,s.
The total SG stored energy is
$E_\mathrm{SG} = H_1 S_1 + H_2 S_2 = 2{,}462$\,MVA$\cdot$s.
Since the event is a pure load step ($V_g \approx 1.0$\,pu, no VRT
activation), the power bound~\eqref{eq:Hpower} evaluates to
$\HP = (\kappa - \lambda)\,f_0 / (2\,\mathrm{RoCoF}_\mathrm{max})$.
The power bound becomes binding only when $\HP < \HC$, i.e.\
when $\lambda > \lambda^{*} = \kappa - 2\HC\,\mathrm{RoCoF}_\mathrm{max}/f_0$.
For the present system, $\lambda^{*} = 1.2 - 2(4.7)(1.64)/50 = 1.012 > 1$,
so $\HP > \HC$ for \emph{all} feasible loading levels
$\lambda \in [0,\,1]$.
The envelope~\eqref{eq:envelope} therefore predicts
$\Heff = \HC = 4.7$\,s regardless of the dispatch point, provided
$\HE$ is non-binding.
The top panels of Fig.~\ref{fig:sim_results} present the unfiltered
frequency and instantaneous RoCoF during the first 150\,ms.
In the DC-link-only case the initial RoCoF is $-1.64$\,Hz/s;
with ESS this reduces to $-1.33$\,Hz/s.
To verify that the $K_\mathrm{inertia}$ from the connection agreement
is reflected in the measured RoCoF, the DC-link case
($\HE \approx 0.5$\,ms, no sustained inertial contribution) serves
as the SG-only reference.
The swing equation relates the initial RoCoF to the total system
stored energy:
\begin{equation}
  \left\lvert\frac{df}{dt}\right\rvert_{t=0^+}
  = \frac{\Delta P\,f_0}{2\,E_\mathrm{sys}}.
  \label{eq:swing_rocof}
\end{equation}
In the DC-link case, the inverter cannot sustain inertial injection,
so $E_\mathrm{sys}^\mathrm{(DC)} \approx E_\mathrm{SG}$.
In the ESS case, the inverter contributes an additional
$H_\mathrm{inv}\,S_\mathrm{inv}$, giving
$E_\mathrm{sys}^\mathrm{(ESS)} = E_\mathrm{SG} + H_\mathrm{inv}\,S_\mathrm{inv}$.
Since the disturbance and network are identical, dividing
\eqref{eq:swing_rocof} for the two cases yields:
\begin{equation}
  \frac{\lvert\mathrm{RoCoF}_\mathrm{DC}\rvert}
       {\lvert\mathrm{RoCoF}_\mathrm{ESS}\rvert}
  = \frac{E_\mathrm{SG} + H_\mathrm{inv}\,S_\mathrm{inv}}
         {E_\mathrm{SG}}.
  \label{eq:rocof_ratio}
\end{equation}
Solving for the inverter inertia:
\begin{equation}
  H_\mathrm{inv}^\mathrm{meas}
  = \frac{E_\mathrm{SG}}{S_\mathrm{inv}}
    \!\left(
      \frac{\lvert\mathrm{RoCoF}_\mathrm{DC}\rvert}
           {\lvert\mathrm{RoCoF}_\mathrm{ESS}\rvert} - 1
    \right).
  \label{eq:H_meas_formula}
\end{equation}
Substituting the measured values:
\begin{equation}
  H_\mathrm{GFL}^\mathrm{meas}
  = \frac{2{,}462}{125}
    \!\left(\frac{1.64}{1.33} - 1\right)
  \approx 4.6~\mathrm{s},
  \label{eq:H_meas}
\end{equation}
which is within 2\% of the framework prediction
$\Heff = \HC = 4.7$\,s.
The RoCoF reduction from $-1.64$ to $-1.33$\,Hz/s therefore confirms
that the inverter's $K_\mathrm{inertia}$ is fully delivered to the
system when backed by adequate collocated storage.
\subsection{VRT Impact on Inertia Delivery}
\label{sec:sim_vrt}
The previous subsection demonstrated that, in the absence of a significant
voltage disturbance ($V_g \approx 1.0$\,pu), the GFL inverter delivers
its full control-limited inertia $\HC$.
This test examines what happens when VRT is simultaneously requested.
A permanent voltage depression to $V_g = 0.62$\,pu is applied at
Bus~10 with reactive-current gain $K_q = 5$.
The requested VRT reactive current from~\eqref{eq:IqVRT} is
$I_{q,\mathrm{VRT}} = K_q (V_\mathrm{th} - V_g) / V_\mathrm{nom}
 = 5 \times (0.9 - 0.62) / 1.0 = 1.4$\,pu.
Combined with the pre-event reactive current, the total reactive
demand exceeds $\kappa = 1.2$\,pu, leaving
$I_{p,\mathrm{avail}} \approx 0$ from~\eqref{eq:Ipavail} and hence
$\kappa_\mathrm{eff} \approx 0$ from~\eqref{eq:kappaeff}.
The power bound~\eqref{eq:HpowerVRT} then gives
$\HP^\mathrm{VRT} \leq 0$ for any loading $\lambda$, and the envelope
\eqref{eq:envelope} reduces to $\Heff = 0$: the inverter cannot
contribute any sustained inertia.
The bottom panels of Fig.~\ref{fig:sim_results} confirm this
prediction.
Under the same load step, the GFL~(VRT) case exhibits a peak RoCoF
of $-1.79$\,Hz/s compared to $-1.33$\,Hz/s for GFL~(no~VRT).
Not only does the inverter fail to provide inertia; the VRT-case
RoCoF ($-1.79$\,Hz/s) is slightly worse than the SG-only reference
($-1.64$\,Hz/s from the DC-link case in Section~\ref{sec:sim_ess}).
This is because the PLL dynamics under the depressed voltage cause a
transient reduction in active-power delivery during the initial
measurement window, momentarily degrading the frequency response
beyond the SG-only baseline.
The degradation is modest in magnitude (${\approx}9$\%), but it is
physically meaningful: VRT activation can actually worsen---not merely
suppress---the inverter's contribution to RoCoF.
\section{Operational Applications}
\label{sec:operational}
\subsection{Guaranteed Inertia for System Planning}
From the known dispatch $\lambda_i$ and a design-basis voltage
$V_{g,\min}^\mathrm{design}$, the TSO computes per-plant guaranteed
inertia $H_i^\mathrm{guar} = \Heff(\lambda_i, V_{g,\min}^\mathrm{design})$
and aggregates:
\begin{equation}
  H_\mathrm{sys}^\mathrm{guar}
  = \sum_{j\in\mathrm{SG}} H_j
  + \sum_{i\in\mathrm{IBR}} \frac{H_i^\mathrm{guar}\,\Srat^{(i)}}
                                  {S_\mathrm{sys}}
  \label{eq:Hguaranteed}
\end{equation}
This is directly analogous to the existing SG inertia sum, but
IBR contributions are now \emph{dispatch-dependent} and
\emph{voltage-contingency-conditioned}---a finding consistent with
recent AEMO analysis showing that GFM BESS synthetic inertia is
variable and depends on operating point, contingency size, and
overload capacity \cite{AEMO2024GFM}.
Real-world validation of GFM (``virtual machine mode'') inertia
delivery at utility scale from a 150\,MW BESS is reported in
\cite{ARENA2022VMM}, providing operational support for the envelope
described here.
\subsection{IBR Inertia Capability Curve}
We propose the \emph{inertia capability curve} $H(\lambda, V_{g,\min})$
as a grid-code instrument. The curve:
\begin{itemize}
  \item is derived analytically from the framework (or validated
    through type testing per \cite{EentsoEgridcode2022,IEEE2800});
  \item specifies guaranteed $H$ as a function of $\lambda$ and
    design-basis $V_{g,\min}$;
  \item is used as part of the grid connection agreement; and
  \item enables IBR inertia to be counted in planning studies with
    the same rigor as SG inertia.
\end{itemize}
For SGs, the equivalent curve reduces to a constant,
$H = \mathrm{const}$; for IBRs, both the loading and the design-basis
voltage dimension are required.
\subsection{Complementarity with Online Inertia Estimation}
Online real-time estimators---e.g.\ the data-driven IBR estimator of \cite{Tan2023}---estimate the
current system inertia and provide valuable
situational awareness. However, they are:
(i)~instantaneous or backward-looking;
(ii)~blind to the constraint that causes the observed value; and
(iii)~subject to the RoCoF measurement-window effect derived in
Section~\ref{sec:control}.
The proposed bounds framework is \emph{complementary}: it provides
the worst-case guaranteed system inertia and identifies the
constraint that sets it.
Deviations between the online estimate $\hat{H}_\mathrm{sys}(t)$ and
the analytically predicted $H_\mathrm{sys}^\mathrm{guar}(t)$ serve as
an anomaly signal, indicating unmodeled dynamics, model errors, or
estimation artifacts.
This parallels voltage-stability practice, where TSOs combine
real-time voltage monitoring with offline P--V curve analysis that
maps the stability boundary. Without the bounds framework, an operator
would know the current inertia level but not whether that level is
robust to a credible voltage depression.
\section{Conclusions}
\label{sec:conclusions}
A unified analytical framework for the achievable inertia constant of
inverter-based resources has been presented, expressed as
$\Heff = \min(\HE, \HP, \HC)$ and validated on the IEEE 9-bus system.
The following conclusions are drawn.
\textbf{1.} Achievable IBR inertia varies with loading, grid voltage,
and time after the disturbance. The inertia envelope
\eqref{eq:envelope} quantifies this dependence using
TSO-accessible plant parameters, in contrast to synchronous
generators, where inertia is a fixed machine constant.
\textbf{2.} During disturbances that produce a voltage depression,
VRT reactive-current priority can reduce the active-power headroom
to zero when $V_g$ falls below a loading-dependent critical voltage
$V_g^*$.
This constraint has no equivalent in synchronous generators and is
not captured by single-constraint inertia analyses.
\textbf{3.} GFL plants provide no inertia at the instant of a
disturbance, and their contribution ramps up to its full value only
over the activation delay $\tact \approx 100$--$200$\,ms.
The closed-form expression derived in Section~\ref{sec:control}
shows that this activation delay is the primary source of the
window-length dependence observed in TSO inertia estimates.
\textbf{4.} For GFL plants without collocated ESS,
each 1\% of available power headroom below the maximum power point adds
approximately 0.25\,s of achievable inertia at 50\,Hz and 1\,Hz/s
RoCoF, up to the control-limit crossover at $\lambda \approx 0.76$.
\textbf{5.} The two-dimensional capability curve $H(\lambda, V_{g,\min})$
provides a format for grid connection agreements that enables IBR
inertia to be assessed in planning studies with the same methodology
as synchronous generator inertia.
Summing the $H$ values from connection agreements of online IBR
plants, without accounting for operating-state dependence, may
overestimate effective inertia under adverse but credible conditions.
\balance
\bibliographystyle{IEEEtran}
\bibliography{bibliography}
\end{document}